\documentclass[prl,floatfix,reprint,twocolumn,preprintnumbers,amsmath,amssymb,showpacs,superscriptaddress]{revtex4}
\usepackage[T1]{fontenc}
\usepackage[english]{babel}
\usepackage[latin9]{inputenc}
\usepackage{amstext}
\usepackage{graphicx}
\usepackage{esint}
\usepackage{float}
\usepackage{placeins}
\usepackage{hyperref}
\hypersetup{colorlinks=true, citecolor=blue, urlcolor=blue, linkcolor=blue}

\usepackage{color}
\usepackage{float}
\usepackage{amsmath}
\usepackage{amssymb}
\PassOptionsToPackage{normalem}{ulem}
\usepackage{ulem}



\begin{document}
	
\title{Continuum approach to real time dynamics of 1+1D gauge field theory:\\
out of horizon correlations of the Schwinger model}
\author{Ivan Kukuljan}
\affiliation{Max-Planck-Institute of Quantum Optics, Hans-Kopfermann-Str. 1, DE-85748 Garching, Germany}
\affiliation{Munich Center for Quantum Science and Technology, Schellingstr. 4, DE-80799 München, Germany}

\begin{abstract}
We develop a truncated Hamiltonian method to study nonequilibrium
real time dynamics in the Schwinger model - the
quantum electrodynamics in D=1+1. This is a purely continuum method that captures reliably the invariance under local and global gauge transformations
and does not require a discretisation of space-time. We use it to study a phenomenon that is expected not to be tractable using lattice methods: we show that the 1+1D quantum electrodynamics admits the dynamical horizon violation effect which was recently discovered in the case of the sine-Gordon model.  Following a quench of the model, oscillatory long-range correlations develop, manifestly violating the horizon bound. We find that the oscillation frequencies of the out-of-horizon correlations correspond to twice the masses of the mesons of the model suggesting that the effect is mediated through correlated meson pairs. We also report on the cluster violation in the massive version of the model, previously known in the massless Schwinger model. The results presented here reveal a novel nonequilibrium phenomenon in 1+1D quantum electrodynamics and make a first step towards establishing that the horizon violation effect is present in gauge field theory.
\end{abstract}


\maketitle

\emph{Introduction. - }
Computing real time dynamics of an interacting many-body quantum system
is a notoriously difficult problem. It has been currently getting an overwhelming amount of attention due to the fast developing
field of nonequilibrium physics both in high energy \cite{Kamenev2011,Berges2004a,Berges2004b,Calzetta2008,Grozdanov2015,Grozdanov2015b,CalabreseCardyQuenchesReview2016,Doyon2016,Glorioso2018}
and condensed matter physics \cite{Giamarchi2015,Moore2016,Medenjak2017}
on one side and renewed
interest in chaos and information scrambling on the other side \cite{Susskind2008,Kitaev2014,Maldacena2016,Polchinski2016,Jahnke2019}.
It is also becoming a matter of increased experimental importance
\cite{SchmiedmayerReview2015,Bloch2008,Lukin2017,Madan2018}
. The set of tools to deal with the problem has been
greatly enriched by developments and new insights in integrability
theory \cite{LeClairMussardo1999,Essler2016,Caux2016}, holography
\cite{Maldacena1999,Aharony2000,CasalderreySolana2014,Zaanen2015,Liu2018}
and numerical algorithms such as density matrix renormalisation group (DMRG) \cite{White1992,Schollwoeck2011}, tensor networks (TNS) \cite{Cirac2009,Orus2014,Bridgeman2017} and  lattice gauge theory \cite{Emonts2020A,Emonts2020B}. Although
in the present time, there is an abundance of excellent numerical
methods available for discrete systems, the methods for the real time
evolution directly in the continuum remain scarce and less developed.

A powerful class of algorithms are the \textit{truncated Hamiltonian
methods} (THM) \cite{YurovZomolodchikovTCSA1990,KonikReviewNonperturbative2018,YurovFermionicTCSA1991,MussardoTCSA1991,TakacsTCSAc1998,TakacsTCSAsGneumann2001,TakacsTCSAsGdirichlet2002,Rychkov2015,Rychkov2018}.
They are numerical methods for \textit{quantum field theories} (QFT) that work in the continuum and do not require a discretisation of space-time.
They can be applied to a wide set of tasks like computing spectra
\cite{YurovZomolodchikovTCSA1990,YurovFermionicTCSA1991,MussardoTCSA1991,TakacsTCSAc1998,TakacsTCSAsGneumann2001,TakacsTCSAsGdirichlet2002,Rychkov2015,Rychkov2017,Rychkov2018,Miro2020} and level spacing statistics \cite{Brandino2010,Srdinsek2020}, studying symmetry breaking
\cite{Rychkov2018}, correlation
functions \cite{KukuljanCorrelations2018,Kukuljan2020}, real time
dynamics \cite{TakacsIsingQuenches2016,KukuljanCorrelations2018,TakacsIsingQuenches2018,Horvath2018,Kukuljan2020}
and also gauge field theories \cite{TakacsWZW2015,TakacsQCD2016}.
The class of methods originates from the \textit{truncated conformal
space approach} (TCSA) introduced by Yurov and Zamolodchikov
\cite{YurovZomolodchikovTCSA1990}. A QFT model on a compact domain is regarded as point
along the \textit{renormalisation group} (RG) flow from the \textit{ultra violet} (UV) fixed point generated by a relevant
perturbation. The conformal field theory (CFT) algebraic machinery is used to represent the
Hamiltonian as a matrix in the basis of the UV fixed point CFT Hilbert space. Finally, an energy cutoff is
introduced to obtain a finite matrix which enables numerical computation that indeed efficiently captures nonperturbative effects.
More broadly, instead of CFT, any solvable QFT can be used as the
starting point for the expansion.

One of the central properties of quantum physics out of equilibrium
is the \textit{horizon effect} introduced by Cardy and Calabrese  \cite{CalabreseCardyQuenches2006,CalabreseCardyQuenches2007,Igloi2000}.
A quantum system is initially prepared in a short range correlated
nonequilibrium state, $\left\langle O(x)O(y)\right\rangle\propto e^{-\left|x-y\right|/\xi}$
with  a local observable $O$, the correlation length $\xi$, and let to evolve
dynamically for $t>0$ - a protocol commonly termed a \textit{quantum
quench}. The horizon bound states that the connected correlations following the quench
spread within the horizon: $\left|\left\langle O(t,x)O(t,y)\right\rangle\right|<\kappa \,e^{-\text{max}\left\{ (\left|x-y\right|-2ct)/\xi_{h},0\right\} }$ for some constant $\kappa$,
where $\xi_{h}$ is called the \textit{horizon thickness} and $c$
is the maximal velocity of the theory - speed of light in QFT and the Lieb-Robinson (LR) velocity in discrete systems \cite{LiebRobinson1972}.
The intuition is that correlations spread by pairs
of entangled particles created in initially correlated region
$\left|x-y\right|\lesssim\xi$ and traveling to opposite directions. This bound has been rigorously
proven in CFT \textit{\cite{CalabreseCardyQuenches2006,CalabreseCardyQuenches2007,Cardy2016}}
and demonstrated, analytically and numerically in a large set
of interacting systems
\cite{CalabreseCardy2005,CalabreseChiara2006,Burrell2007,FagottiCalabrese2008,Laeuchli2008,Eisler2008,Manmana2009,CalabreseTransverseIsing2011,Igloi2012,CalabreseTransverseIsingI2012,CalabreseTransverseIsingII2012,EsslerLocalQuenches2012,EsslerQuench2012,BardarsonMoore2012,Kim2013,Hauke2013,Schachenmayer2013,Richerme2014,Carleo2014,Nezhadhaghighi2014,BonnesEssler2014,ColluraCalabrese2014,Krutitsky2014,CalabreseQuenchExcitedStates2014,CalabreseQuenchExcitedStatesII2014,Altman2014,SotiriadisLongRange2015,BuyskikhEssler2016,Altman2015,FagottiPrethermalisation2015,BertiniLightcone2016,Cardy2016,DoyonHydro2016,BertiniHydro2016,BertiniLightcone2016,Zhao2016,Pitsios2017,TakacsConfinement2017}   as well as observed in experiments \cite{Cheneau2012,Jurcevic2014,Langen2013}.
It has therefore been believed to be a universal
property of quantum physics. 

\begin{figure}[htbp]
	\begin{centering}
		\includegraphics[width=1\columnwidth]{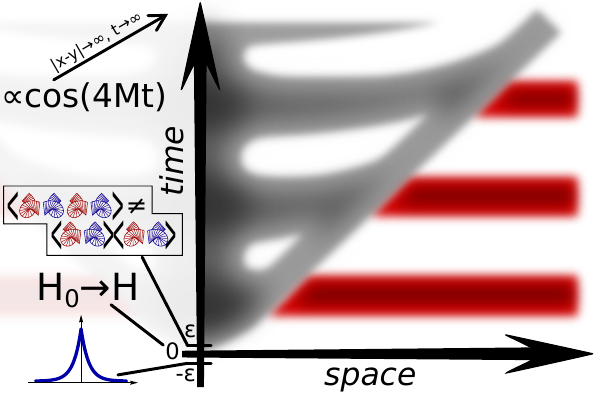}
		\par\end{centering}
	\caption{Dynamical horizon violation as found in the sine-Gordon model \cite{Kukuljan2020}. The system is prepared in the ground state of a gaped Hamiltonian $H_0$ with short range correlations $\propto e^{-\left|x-y\right|/\xi}$. At time $t=0$ the Hamiltonian is quenched to $H$. This generates cluster violating 4-point correlations of solitons and antisolitons, eq. \eqref{eq:ClusterViolation} (here symbolically pictured using classical solitons), which are not observable at $t=0$ but result in oscillating out-of-horizon correlations of local observables $\left\langle O(-x)O(x)\right\rangle$ at later times. The horizon is depicted here with gray color and the horizon violating correlations with red. Asymptotically, the latter oscillate with a frequency 4 times the soliton mass $M$, respectively twice the breather masses in the attractive regime.  \label{fig:ViolationSG}}
\end{figure}

In a recent publication together with Sotiriadis and Tak{\'{a}}cs
\cite{Kukuljan2020}, we have demonstrated that the horizon bound
can be violated in QFT with nontrivial topological properties. We
have proved this in the case of the \textit{sine-Gordon} (SG) field theory,
a prototypical example of strongly correlated QFT
\begin{equation}
\mathcal{L}_{SG}=\frac{1}{2}(\partial_\mu\Phi)(\partial^\mu\Phi)+\frac{\mu^{2}}{\beta^{2}}\cos(\beta\Phi)\label{eq:SGhamiltonian}
\end{equation}
Starting from short
range correlated states, SG dynamics within a short time generates
infinite range correlations oscillating in time and clearly violating
the horizon bound. The mechanism is the following: Quenches
in the SG model create cluster violating four-body correlations between solitons ($S$) and
anti-solitons ($A$), the topological excitations of the theory, written schematically:
\begin{align}
&\lim\limits_{|x-y|\rightarrow\infty}\left<A(x)S(x+a)A(y)S(y+b)\right>\nonumber\\
&\hspace{2cm} \neq \left<A(x)S(x+a)\right>\left<A(y)S(y+b)\right>\label{eq:ClusterViolation}
\end{align}
The dynamics of the model then converts these solitonic correlations into two-point
correlations of local bosonic fields $\left\langle \Phi(t,x)\Phi(t,y)\right\rangle$,
$\left\langle \Pi(t,x)\Pi(t,y)\right\rangle$ and $\left\langle \partial_{x}\Phi(t,x)\partial_{y}\Phi(t,y)\right\rangle$. There is no violation of relativistic causality involved because the cluster violating correlations \eqref{eq:ClusterViolation} are created by a quench, a global simultaneous event and not by the unitary dynamics of the model which is strictly causal. The
mechanism suggests that the horizon violation
should be found in any QFTs with nontrivial field topologies, an important
class of them being gauge field theories. The results presented in this Letter represent the first steps towards establishing that.

As a consequence of the Lieb-Robinson bound \cite{ProsenLRB2014,BravyLRB2006,LiebRobinson1972}
and the Araki theorem \cite{Araki1969,Eisert2014}, the horizon violation
is expected not to be present in short-range interacting discrete systems with finite local Hilbert space dimension and is likely a
genuinely field theoretical phenomenon. Therefore discretising a model and simulating using DMRG or TNS \cite{KogutSusskind1975,Buyens2015,Buyens2017,Berges2013,Berges2019,ColluraSchwingerConfinement2020,Chanda2020,Magnifico2020,Emonts2020A,Emonts2020B} is not an option so methods working directly in the continuum are needed and THM seem to be the best class of methods for the task.

\emph{The Schwinger model. - } We focus here on the simplest
example of a gauge field theory, the \textit{1+1D quantum electrodynamics} (QED), i.e. the (massive)
\textit{Schwinger model}:
\[
\mathcal{L}=-\frac{1}{4}F_{\mu\nu}F^{\mu\nu}+\bar{\Psi}\left(i\gamma^{\mu}\partial_{\mu}-e\gamma^{\mu}A_{\mu}-m\right)\Psi,
\]
with $\Psi=\left(\Psi_{-},\Psi_{+}\right)^T$ the Dirac fermion, $m$ the electron mass and $e$ the electric charge. 
As a consequence of invariance under large gauge transformations, the model has infinitely degenerate vacuum states, the $\theta$ vacua for a parameter $\theta\in\left[0,2\pi\right)$ that enters the bosonised form of the Hamiltonian and plays the physical role
of the constant background electric field \cite{Coleman1975,Coleman1976}. The Schwinger model thus has two physical parameters,
the ratio $m/e$ and $\theta$.

The massless $m=0$ version of the model was solved exactly by Schwinger \cite{Schwinger1962} and has a gap of $e/\sqrt{\pi}$
corresponding to a meson, a bound state of a fermion and an antifermion. The full massive
$m>0$ version of the model is not integrable and has a rich phase diagram where the number of mesons depends on the values
of the parameters $m/e$ and $\theta$ \cite{Coleman1975,Coleman1976,KogutSusskind1975,Susskind1976,Crewther1980,Hamer1982,Adam1997,Gutsfeld1999,Gattringer1999,Hamer2000,Giusti2001,Byrnes2002,Christian2006,Cichy2013,Banuls2013,Buyens2015,BuyensThesis2016}. The Schwinger
model displays confinement and has been extensively studied for pair creation and string breaking
\cite{Coleman1975,Nakanishi1978,Nakawaki1980,Gross1996,Hosotani1996,Cooper2006,Chu2010,Berges2013,Klco2018,Zache2019,Berges2019,ColluraSchwingerConfinement2020,Chanda2020,Magnifico2020,McGady2020}.

Finally, it is known that due to the vacuum degeneracy,
the massless version of the Schwinger model exhibits cluster violation of correlators of chiral fermion densities $\rho_{\pm}(x)=N\left[\bar{\psi}(x)\frac{1\pm\gamma^{5}}{2}\psi(x)\right]$,
\cite{Lowenstein1971,Montalbano1994,Abdalla2001},
\begin{equation}
\left\langle \rho_{-}(x_{1})\cdots\rho_{-}(x_{n})\rho_{+}(y_{1})\cdots\rho_{+}(y_{n})\right\rangle,\label{eq:ChiralDensities}
\end{equation}
closely related to the correlators from eq. \eqref{eq:ClusterViolation}.  This makes the model a good candidate for
the horizon violation. The cluster violation is also intimately related to confinement of gauge theories \cite{LowdonClusterViolation2016,Lowdon2017,Lowdon2018,LowdonAnalyticStructureQCD2018}.

Here we study general quenches
of the massive Schwinger model and focus on the spreading of the current-current
correlators:
\begin{align}
C_{\mu}(t,x,y) & =\left\langle J^{\mu}(t,x)J^{\mu}(t,y)\right\rangle .\label{eq:Correlator}
\end{align}
We prepare the system in the ground state
of the model with the prequench values of the parameters $m_{0}/e_{0}$,
$\theta_{0}$ and at time $t=0$ switch the parameters to their postquench
values $m/e$, $\theta$.

\emph{The method. - }
We implement a THM for
the Schwinger model in finite volume $L$ with anti-periodic boundary conditions (Neveu-Schwarz sector). We elimination the gauge redundancy of degrees of freedom alongside with the bosonisation of the model \cite{IsoMurayama1990}.

Choosing the
Weyl (time) gauge, $A_{t}=0$, and defining $A\equiv A_{x}$, the Hamiltonian of the model is
$
H=\int_{0}^{L} dx\left(\frac{1}{2}\dot{A}^{2}-\bar{\Psi}\left[\gamma^{1}(i\partial_{x}-eA)-m\right]\Psi\right).
$
Expanding the fermion currents
$
J_{\sigma}(x)=\Psi_{\sigma}^{\dagger}(x)\Psi_{\sigma}(x)=\frac{1}{L}\left[Q_{\sigma}-\sigma\sum_{n>0}\sqrt{n}\left(b_{\sigma,n}e^{-\sigma in\frac{2\pi}{L}x}+b_{\sigma,n}^{\dagger}e^{\sigma in\frac{2\pi}{L}x}\right)\right]$, with the chirality $\sigma=\pm$,
its modes obey bosonic canonical commutation relations. Further defining the $N_\sigma$ vacua as 
$
\left|0;N_{-}\right\rangle\equiv\prod_{n=N_{-}}^{\infty}c_{-,n}^{\dagger}\left|0\right\rangle$, $\left|0;N_{+}\right\rangle\equiv\prod_{n=-\infty}^{N_{+}-1}c_{+,n}^{\dagger}\left|0\right\rangle 
$, with $c_{\sigma,n}$ the fermion mode operators,
the Hilbert space spanned by bosonic modes $b_{\sigma,n}^{\dagger}$ on top of $\left|0;N_{-}\right\rangle\otimes\left|0;N_{+}\right\rangle$ is equivalent to the Hilbert space spanned by  $c_{\sigma,n}^{\dagger}$ acting on top of $\left|0\right\rangle$. This is the foundation for the bosonisation of the model. Because of the invariance under large gauge transformations, the true vacua of the system are the infinitely degenerate $\theta$ vacua
$
\left|\theta\right\rangle =\sum_{N\in\mathbb{Z}}e^{-iN\theta}\left|0;N\right\rangle
$
for $\theta\in[0,2\pi)$. Gauge invariance  further implies that the only mode of the EM potential $A$ that is not fixed by the Gauss law is the zero mode $\alpha=\frac{1}{L}\int_{0}^{L}dx\,A(x)$ along with its its dual $i\partial_\alpha=\int_{0}^{L}dx\,\dot{A}(x)$.

By setting $B_0=\sqrt{\frac{1}{2ML}}\left(-\sqrt{\pi}\left\{Q_+-Q_-\right\}+\frac{\partial}{\partial\alpha}\right)$, the part of the Hamiltonian involving the zero modes transforms into a harmonic oscillator with the mass
$
M=\frac{e}{\sqrt{\pi}}
$. 
Complemented with a Bogoliubov transform of the nonzero momentum modes into massive bosonic modes:
$B_{\sigma,n}=\frac{1}{2}\left(\frac{\sqrt{E_{n}}}{\sqrt{k_{n}}}+\frac{\sqrt{k_{n}}}{\sqrt{E_{n}}}\right) b_{\sigma,n} -\frac{1}{2}\left(\frac{\sqrt{E_{n}}}{\sqrt{k_{n}}}-\frac{\sqrt{k_{n}}}{\sqrt{E_{n}}}\right) b_{-\sigma,n}^{\dagger}$, with $k_n=\frac{2\pi n}{L}$ and $E_n=\sqrt{M^2+k_n^2}$, the massless part of the Hamiltonian is transformed into the Hamiltonian of a massive free boson with the mass $M$. The mass term of the Hamiltonian is written in the bosonic form using the bosonisation relation
$
\Psi_{\sigma}(x)=F_{\sigma}\frac{1}{\sqrt{L}}:\negmedspace e^{-\sigma i \left(\sqrt{4\pi}\Phi_{\sigma}(x)-\frac{\pi}{L} x \right)}\negmedspace:$
with $\partial_{x}\Phi_{\sigma}(x)=\sqrt{\pi}J_{\sigma}(x)+\frac{\sigma e}{2\sqrt{\pi}}\,A(x)$ the chiral boson field and $F_{\sigma}$ the Klein factor. Then using $F_{\sigma}^{\dagger}F_{-\sigma}\left|\theta\right\rangle=e^{\sigma i\theta}\left|\theta\right\rangle$, the Schwinger model Hamiltonian takes the bosonised form
\begin{align}
H & =H_{M}+U,\label{eq:THMhamiltonianSchwinger} \\
H_{M} &=M\left(B_{0}^{\dagger}B_{0}\right)+\sum_{n>0}E_{n}\left(B_{+,n}^{\dagger}B_{+,n}+B_{-,n}^{\dagger}B_{-,n}\right),\nonumber\\
U&  =-\frac{mM}{2\pi}e^{\gamma}\int_{0}^{L}dx\,:\negmedspace\cos\left(\sqrt{4\pi}\Phi(x)+\theta\right)\negmedspace:_{M}
.\nonumber
\end{align}
with $\Phi(x)=\Phi_- + \Phi_+$ with Bogoliubov transformed modes, $:\bullet:_{M}$ denotes normal ordering w.r.t. the mass $M$ and $\gamma$ is the Euler-Mascheroni constant. 

The form of the Hamiltonian \eqref{eq:THMhamiltonianSchwinger} offers a natural THM splitting into the massive free part and the cosine potential. To implement the numerical method, the cosine potential and the observables, are
represented as matrices in the Hilbert space of the free part - the Fock space generated by applying the $B_{\sigma,n}^\dagger$ modes on the $\theta$ vacuum. Finally, an energy cutoff $\left\langle \Psi\right|H_{M}\left|\Psi\right\rangle\leq E_{\text{cut}}$
is imposed on the states $\left|\Psi\right\rangle$ of the THM Hilbert space.  Momentum conservation implied by translation invariance and the decoupling of the $B_0$ mode from the rest of the modes are used to further reduce the dimension of the Hilbert space by diagonalising each sector separately. We use the Hilbert spaces with up to 20 000 states per sector. The full details of the method can be found in the Supplemental Material \cite{SupplementaryMaterial}.

\begin{figure}[htbp] 
	\begin{centering}
		\includegraphics[width=1\columnwidth]{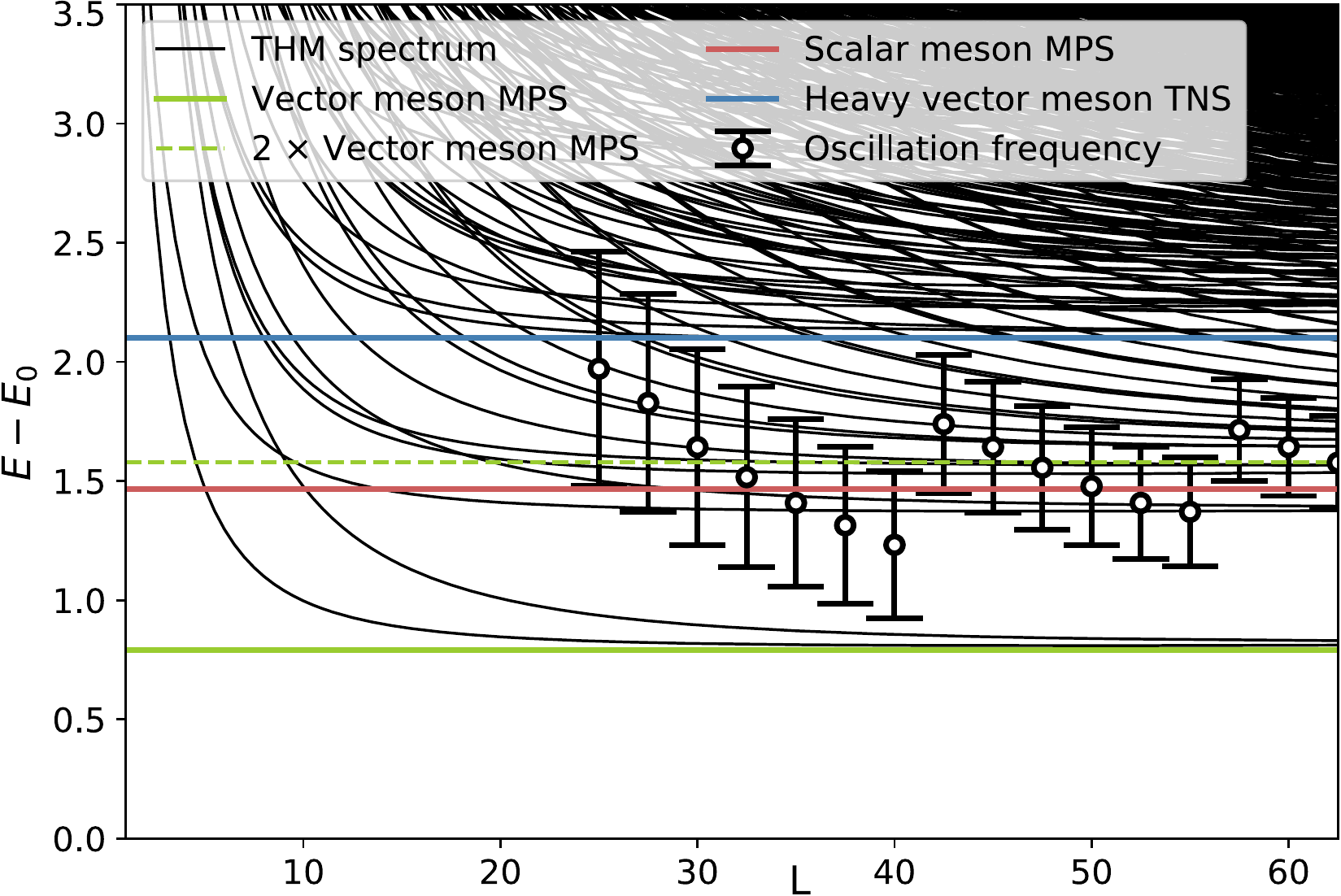}
		\par\end{centering}
	\caption{The THM spectrum the Schwinger model at $m/e=0.125$ in dependence of
		the system size $L$ in the 0, 1 and 2 sectors of the total momentum.
		The spectral lines are compared with the $L\rightarrow\infty$ results of the  MPS computations \cite{Banuls2013} for the vector and the
		scalar particles and the TNS \cite{BuyensThesis2016}
		for the heavy vector particle. On top of the spectrum, the dominant frequency of the oscillations of the out-of-horizon correlations are plotted. \label{fig:Spectrum}}
\end{figure}

\begin{figure*}[htbp]
	\begin{centering}
		\includegraphics[width=\textwidth]{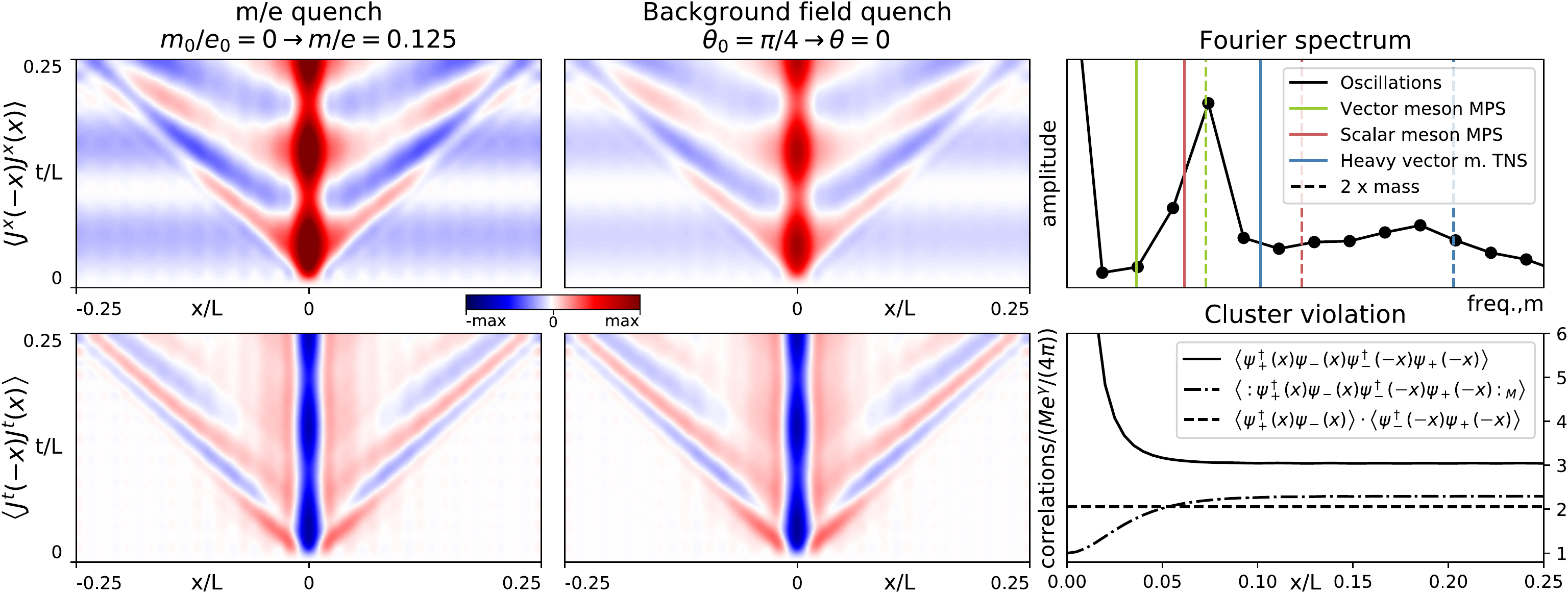}
		\par\end{centering}
	\caption{Left: Time dependent $\left\langle J^{x}(t,x)J^{x}(t,y)\right\rangle $
		and $\left\langle J^{t}(t,x)J^{t}(t,y)\right\rangle $ correlations
		for different type of quenches in the Schwinger model (initial correlations subtracted): 1.) Quench in $m/e$ with $m_{0}=0$,
		$m=0.125$, $\theta_{0}=\theta=0$; 2.) Quench
		in $\theta$ with $\theta_{0}=\frac{\pi}{4}$, $\theta=0$, $m_{0}=m=0.125$. Both with $e_{0}=e=1$, $L=40$. Upper right: Frequency spectrum of the out-of-horizon component of the correlations (mass quenches to $m=0.25$, $e=1$, $L=47.5$) compared to meson masses (full lines) and twice the values of meson masses (dashed lines). Lower right: Cluster violation in the massive Schwinger model at $m=0.125$, $\theta=0$, $e=1$, $L=40$.
		\label{fig:SchwingerQuenches}}
\end{figure*}

\emph{Results. - }
Our THM implementation of the Schwinger model recovers the results from the literature for
the meson masses and gives a region of highly dense states above them, referred to as the continuum in the $L\rightarrow\infty$ limit (fig. \ref{fig:Spectrum}). This serves as a sanity check of the method. We are able to get the masses of the vector meson precisely, while our THM method seems to be slightly less precise for the scalar meson mass.  We have been able to simulate large system sizes $L\gg \frac{1}{M}$ where the finite size effects are exponentially suppressed.

The results shown in fig. \ref{fig:SchwingerQuenches} indeed
confirm that the Schwinger model exhibits the horizon violation effect - the correlation functions
$C_{x}(t,x,y)$ are
nonzero and oscillating for $|x-y|>2t$. The effect is found in quenches in both $e/m$ and $\theta$ as well as in quenches to and from the massless Schwinger model. The sign of the out-of-horizon correlations changes depending whether the quenched parameter is increased or decreased. As
is expected for periodic boundary conditions, the effect is present
in the $C_{x}$ and not present in the $C_{t}$ channel. 

To shed light on the origin of the effect, we study the clustering properties of correlators of chiral densities \eqref{eq:ChiralDensities} (fig. \ref{fig:SchwingerQuenches}, lower right), more specifically, its component $\left\langle\psi_\sigma^\dagger(x)\psi_{-\sigma}(x)\psi_{-\sigma}^\dagger(y)\psi_{\sigma}(y)\right\rangle$. We find that the correlator violates clustering - when $x$ and $y$ are far apart, the correlator does not cluster into $\left\langle\psi_\sigma^\dagger(x)\psi_{-\sigma}(x)\right\rangle\left\langle\psi_{-\sigma}^\dagger(y)\psi_{\sigma}(y)\right\rangle$. In case of the massless Schwinger model, this clustering violation is well known  and can be computed analytically \cite{Lowenstein1971,Montalbano1994,Abdalla2001}, in case of the massive version of the model, this is to our knowledge a new result. Interestingly, in the massless case, the normal ordered version of the correlator does not exhibit the clustering violation while in the massive model, even the normal ordered correlator violates clustering. We expect that similarly as in the SG model \cite{Kukuljan2020}, the nonlinear postquench dynamics rotates the initial clustering violation from such chiral correlators into the local nonchiral observables. We note that in case of the ground states of the massive model, we observe numerically a tiny clustering violation also in the $C_x$ correlators which is two orders of magnitude smaller than the cluster violation of $\left\langle\rho_\sigma\rho_{-\sigma}\right\rangle$. We expect, however, that this is not a physical fact but an artifact of the THM truncation. Such tiny artifacts are common in derivative fields but do not falsely produce the horizon violation effect, as was for example verified in case of Klein-Gordon dynamics in the first version of \cite{Kukuljan2020,Kukuljan2020v1}. As well as that, our THM simulation of the Schwinger model displays the horizon violation in the quenches starting from the massless model, where there are no such artifacts in the initial state. So we expect that the effect originates fully from the cluster violation of the chiral terms.

Fig. \ref{fig:Spectrum} shows how the dominant frequencies of the oscillations compare to the spectrum of the model. Due simulation times limited to $t\leq L/4$, we are only able to see a few oscillations. Therefore, the frequencies have considerable error bars ($\Delta\omega\approx 2\pi/L$ - half a frequency bin) and the values of the possible discrete frequencies move with $L$ resulting in a chainsaw pattern. The error bars compare to both the scalar meson mass and twice the vector meson mass. Based on the mechanism of the effect in the SG model \cite{Kukuljan2020}, it is expected that the frequencies correspond to twice the mass of the lightest meson. This is supported by computations at higher values of $m/e$, where those masses can be better discriminated (Fourier spectrum in the upper right of   fig. \ref{fig:SchwingerQuenches}).  This suggest that the horizon violation is mediated through correlated vector meson pairs entangled by the quench. In some cases even subdominant peaks appear close to twice the masses of heavier mesons in the frequency spectra, suggesting that they could also be contributing to the effect.

\emph{Discussion. - }
We stress again that the observed phenomenon is in no contradiction
with relativistic causality as guaranteed by the Lorentz invariance of the model the micro causality of the fields. Rather, the violation of horizon
can be likely traced back to the cluster violation of chiral fermion fields as in the SG model \cite{Kukuljan2020}. 

Using the simplest representative, we have hereby demonstrated that the horizon violation occurs in gauge field theory. In the future, it  would be interesting
to explore higher gauge  theories like SU(2) or SU(3) or
study the Wess--Zumino--Witten models. It would be of crucial importance
to answer whether the effect is present also in $D>1+1$. There, gauge fields are dynamical, so the physics could be drastically different.
Further analytical approaches should be found to get a better
understanding of the effect in the Schwinger model. 

\begin{figure}[H]
	\begin{centering}
		\includegraphics[width=1\columnwidth]{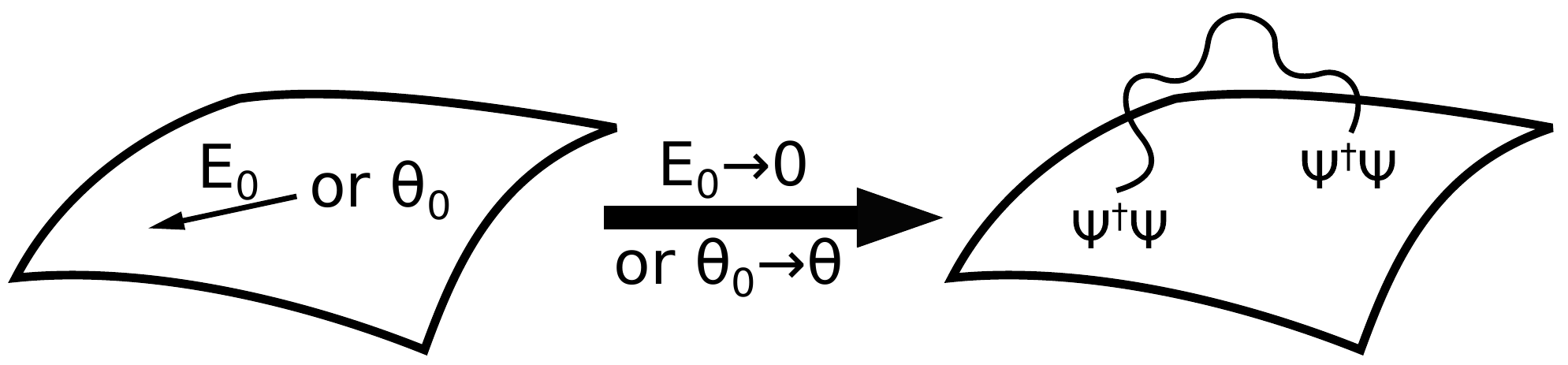}
		\par\end{centering}
	\caption{Decay of the anisotropic initial condition or a $\theta$ term in a toy universe as a quench that generates long range correlations through the horizon violation effect. Long range correla-tions are the price that the toy universe has to pay for the initial anisotropy. \label{fig:CosmologicalQuench}}
\end{figure}

The horizon violation presented in this work is a
novel phenomenon in 1+1D quantum-electrodynamics. It is reasonable to expect that it could have interesting physical implications, in
particular if it turns out that the effect is present also in higher
dimensions. In condensed matter physics, phase transitions are an ubiquitous phenomenon and could serve as a trigger for horizon violation generating quenches. Here, already the $D=1+1$ case could be an interesting
candidate since at the present day there are numerous experiments available for probing 1+1D physics
\cite{Giamarchi2015}. An especially important class are ultra cold
atoms in atom chips, where one dimensional QFTs are directly realised
and correlation functions can be measured both in equilibrium states
and nonequilibrium dynamics \cite{SchmiedmayerCorrelations2017}. In cosmology, there several candidates for quenches like the end of inflation, the QCD and the
electroweak transitions and topological symmetry breaking in grand unified theories \cite{Weinberg2008,BoyanovskyDeVega2006,Gleiser1998,Hindmarsh2020}. Consider also the following example illustrated in fig. \ref{fig:CosmologicalQuench}: a toy universe
is created with an anisotropic initial condition - a nonzero background electric field.
This is a possibility since the zero background field case is a special, fine-tuned, value.
In $D=1+1$ the background electric field is stable while in $D=1+3$, it decays through the electric breakdown of the vacuum \cite{Coleman1976}. The
rapid decay of the background electric field  would serve as a quench that causes a horizon violation effect in the QED degrees of freedom as we have
seen here in the $\theta_{0}\neq0\rightarrow\theta=0$ quenches. This
transforms the initial anisotropy of the toy universe into long range
correlations.  Similarly, in a higher gauge theory the effect could be triggered by a decay of the theta term which is linked in some models with the cosmological constant  \cite{Yokoyama2002,Jaikumar2003}. It would be interesting to explore the possible predictions for traces of this effect in the cosmic microwave background.

Finally, it would be interesting to use THM to
explore the confinement and string breaking phenomena in the Schwinger
model and to use THM implementations \cite{TakacsQCD2016} to study dynamics of higher gauge theories.

\begin{acknowledgements}

This work was supported by the Max-Planck-Harvard Research Center
for Quantum Optics (MPHQ). The author wants to thank Mari Carmen Ba{\~{n}}uls,
Peter Lowdon, Jernej Fesel Kamenik, Miha Nemev\v{s}ek and Sa\v{s}o Grozdanov for useful discussions. Special
thanks to Spyros Sotiriadis for many of our valuable discussions and Gabor Tak{\'{a}}cs for useful discussions and feedback to the first version of the manuscript that helped improve this work.

\end{acknowledgements}

\bibliography{Kukuljan_SchwingerModelHorizon}

\newpage

\onecolumngrid

\FloatBarrier

\

\appendix

\begin{center}
	\Large{\bf{Supplemental Material}}
\end{center}

\section{Details of the THM for the Schwinger model \label{secApp:Details-of-the-method}}

Here we discuss the details of the \textit{truncated Hamiltonian method} (THM)
implementation of the Schwinger model defined on an interval of length $L$ with anti-periodic boundary conditions. 

\subsection{Bosonisation}

The Schwinger model, the quantum electrodynamics in $D=1+1$, is defined by the Lagrangian density
\begin{equation}
\mathcal{L}=-\frac{1}{4}F^{\mu\nu}F_{\mu\nu}+\bar{\Psi}(i\gamma^{\mu}\partial_{\mu}-e\gamma^{\mu}A_{\mu}-m)\Psi
\end{equation}
where $\Psi=\left(\begin{array}{c}
\Psi_{-}\\
\Psi_{+}
\end{array}\right)$
is the Dirac fermion field, $A_{\mu}$ the electromagnetic (EM) potential, $F_{\mu\nu}=\partial_{\mu}A_{\nu}-\partial_{\nu}A_{\mu}$ the EM tensor, $m$ is the electron mass and $e$ the electric charge. Choosing the Weyl (time) gauge, $A_{0}=0$, defining $A\equiv A_x$, the Hamiltonian density takes the form: 
\begin{align}
&\mathcal{H}	=\mathcal{H}_{\text{EM}}+\mathcal{H}_{\text{F}}+\mathcal{H}_{m},\nonumber\\
\mathcal{H}_{\text{EM}}=\frac{1}{2}\dot{A}^{2}, \hspace{1.5cm}	&
\mathcal{H}_{\text{F}}	= -\bar{\Psi}\gamma^{1}(i\partial_{x}-eA)\Psi, \hspace{1.5cm}
\mathcal{H}_{m}	=m\bar{\Psi}\Psi.
\end{align}
We use here the metric $g=\text{diag}(1,-1)$ and the gamma matrices $\gamma^{0}=-\sigma_{1}$, $\gamma^{1}=i\sigma_{2}$.

In order to treat a gauge field theory in Hamiltonian formalism, one has to remove the redundancy in the degrees of freedom coming from gauge invariance. For the Schwinger model this goes hand in hand with bosonisation. Bosonisation
is an exact duality between fermionic and bosonic theories in 1+1D
relativistic QFT, developed by Tomonaga, Mattis,
Lieb,  Mandelstam, Coleman, Haldane and others \cite{Tomonaga1950,MattisLieb1965,SchotteSchotte1969,Mattis1974,LutherPeschel1974,MandelstamSolitons1975,Coleman1975,Haldane1981}. We bosonise the Schwinger model here using the operatorial (constructive) approach of Iso and Murayama \cite{IsoMurayama1990}.

The eigenfunctions and eigenvalues of the massless fermion part of the Hamiltonian $\mathcal{H}_{\text{F}}(x)=\sum_{\sigma=\pm}\sigma \Psi^\dagger(x)(i\partial_{x}-eA)\Psi(x)$ are $(i\partial_{x}-eA)\psi_n=\epsilon_n \psi_n$:
\begin{align}
\psi_{n}(x)&=\frac{1}{\sqrt{L}}e^{-i(\epsilon_{n}x+e\int_{0}^{x}dx\,A)},\nonumber\\
\epsilon_{n}&=\frac{2\pi}{L}\left(n+\frac{1}{2}\delta_{b}-\frac{e\alpha L}{2\pi}\right).\label{eqApp:WaveFuncitons}
\end{align}
The eigenfunctions satisfy periodic boundary conditions (Ramond sector) for $\delta_{b}=0$ and anti-periodic (Neveu-Schwarz sector) for $\delta_{b}=1$. The fermion bosonises to a boson with periodic boundary conditions for both values of  $\delta_{b}$ reflecting the fact that the bosonisation is an equivalence between bosons and fermions up to $Z_2$. We will derive the equations for general $\delta_{b}$ and in the end take $\delta_{b}=1$, the Neveu-Schwarz sector of the fermion, for the THM study. We quantise the fermion field by expansion
\begin{equation}
\Psi(x)=\sum_{n\in\mathbb{Z}}c_{-,n}\psi_{n}(x)\left(\begin{array}{c}
1\\
0
\end{array}\right)+c_{+,n}\psi_{n}(x)\left(\begin{array}{c}
0\\
1
\end{array}\right)
\end{equation}
with the canonical anticommutation relation $\left\{ c_{\sigma,n},c_{\rho,m}^{\dagger}\right\} =\delta_{\sigma,\rho}\delta_{n,m}$. Then $H_{\text{F}}	=H_{+}+H_{-}$ with $H_{\sigma}=\sigma\sum_{n\in\mathbb{Z}}\epsilon_{n}c_{\sigma,n}^{\dagger}c_{\sigma,n}$. 

We expand the EM potential and its conjugate dual as:
\begin{align}
A(x)&=\alpha+\sum_{n\neq0}A_{n}e^{i\frac{2\pi}{L}nx},\nonumber\\
\dot{A}(x)&=i\frac{\delta}{\delta A(x)}	=\frac{i}{L}\left(\frac{\partial}{\partial\alpha}+\sum_{n\neq0}\frac{\delta}{\delta A_{n}}e^{-i\frac{2\pi}{L}nx}\right),\label{eqApp:EMexpansion}
\end{align}
and we shall see that as a consequence of the gauge invariance only the zero modes of the EM field $\alpha$ and $i\frac{\partial}{\partial\alpha}$ are dynamical \cite{Manton1985,IsoMurayama1990}.

Expanding the fermion currents
\begin{equation}
J_{\sigma}(x)=\Psi_{\sigma}^{\dagger}(x)\Psi_{\sigma}(x)=\frac{1}{L}\left[Q_{\sigma}-\sigma\sum_{n>0}\sqrt{n}\left(b_{\sigma,n}e^{-\sigma in\frac{2\pi}{L}x}+b_{\sigma,n}^{\dagger}e^{\sigma in\frac{2\pi}{L}x}\right)\right],\label{eqApp:CurrentExpansion}
\end{equation}
its modes $b_{\sigma,n}=-\frac{\sigma}{\sqrt{n}}\sum_{k\in\mathbb{Z}}c_{\sigma,k}^{\dagger}c_{\sigma,k+\sigma n}$ obey canonical commutation relations $\left[b_{\sigma,n},b_{\rho,m}^{\dagger}\right]=\delta_{\sigma,\rho}\delta_{n,m}$. Further defining the $N_\sigma$ vacua as
\begin{equation}
\left|0;N_{-}\right\rangle\equiv\prod_{n=N_{-}}^{\infty}c_{-,n}^{\dagger}\left|0\right\rangle,\hspace{1.5cm}\left|0;N_{+}\right\rangle\equiv\prod_{n=-\infty}^{N_{+}-1}c_{+,n}^{\dagger}\left|0\right\rangle 
\end{equation}
it can be shown that the Hilbert space spanned by excitations with all possible combinations of $b_{\sigma,n}^{\dagger}$ on top of $\left|0;N_{-}\right\rangle\otimes\left|0;N_{+}\right\rangle$ is equivalent to the Hilbert space spanned by all the possible combinations of $c_{\sigma,n}^{\dagger}$ on top of $\left|0\right\rangle$. This is the core of bosonisation. 

The fermion number operators take the following expectation values on the $N_\sigma$ vacua,
\begin{equation}
\left\langle Q_{\sigma}\right\rangle _{N_{\sigma}}=\sigma\left(N_{\sigma}-\frac{e\alpha L}{2\pi}+\frac{1}{2}(\delta_{b}-1)\right),\label{eqApp:Qexpectations}
\end{equation} 
as can be shown by regularisation by Hurwitz zeta resummation. Similarly, 
\begin{eqnarray}
\left\langle H_{\sigma}\right\rangle _{N_{\sigma}}=\frac{2\pi}{L}\left[\frac{1}{2}\left\langle Q_{\sigma}\right\rangle _{N_{\sigma}}^{2}-\frac{1}{24}\right].\label{eqApp:HexpOnN}
\end{eqnarray}
for $H=\int_{0}^{L}dx\, \mathcal{H}$.

\paragraph{\textbf{Gauge invariance. - }} The fermionic Hilbert space combined with the Hilbert space generated by the modes of the EM modes display a redundancy of degrees of freedom as is characteristic for gauge invariant theories and we have to eliminate this redundancy. QED is invariant under the transformations
\begin{equation}
A_{\mu(x)}\rightarrow A_{\mu}(x)-\partial_{\mu}\lambda(x),\hspace{1.5cm}\psi(x)\rightarrow e^{ie\lambda(x)}\psi(x).
\end{equation}
For systems defined on the circle (and other topologies with nontrivial homotopic groups), the gauge transformations can be divided into \textit{small gauge transformations} where both $\lambda(x)$ and $e^{ie\lambda(x)}$ are single valued and \textit{large gauge transformations} where $e^{ie\lambda(x)}$ is single valued but $\lambda(x)$ is not. Mathematically speaking, small gauge transformations are homotopic to the identity of the Lie group and large gauge transformations are not.

Let's begin with small gauge transformations. As a consequence of the \textit{Dirac conjecture} \cite{GraciaPons1988,KashiwaTakahashi1994} these are represented in the Hilbert space by the operator $U(\lambda)=\exp(-i\int_{0}^{L}dx\,G(x)\lambda(x))$ with the Gauss law generator
\begin{equation}
G(x)=\partial_{x}\left(-i\frac{\delta}{\delta A(x)}\right)-eJ_{0}(x)
\end{equation}
with $J_0=J_+ +J_-$. Requiring that the physical states are invariant under $G$ using the expansions \eqref{eqApp:CurrentExpansion}, \eqref{eqApp:EMexpansion} gives
\begin{align}
Q\left|\text{physical state}\right\rangle &=0\nonumber\\
\left\{ \frac{\delta}{\delta A_{n}}+\frac{eL}{\sqrt{n}2\pi}\left(b_{-,n}^{\dagger}-b_{+,n}\right)\right\} \left|\text{physical state}\right\rangle &=0\nonumber\\
\left\{ \frac{\delta}{\delta A_{-n}}+\frac{eL}{\sqrt{n}2\pi}\left(b_{+,|n|}^{\dagger}-b_{-,|n|}\right)\right\} \left|\text{physical state}\right\rangle &=0\label{eqApp:GaugeInvariance}
\end{align}
For $Q=Q_+ + Q_-$. Taking into account \eqref{eqApp:Qexpectations}, the first constraint means that $N_-=N_+=N$ in physical states and we can define the $N$ vacua as
\begin{equation}
\left|0;N\right\rangle\equiv\left|0;N_{-}\right\rangle\otimes\left|0;N_{+}\right\rangle.
\end{equation}
The second and the third constrain mean that all the nonzero momentum modes of the EM field are fixed by gauge invariance and the only dynamical modes of the EM field are the zero modes $\alpha$ and $i\frac{\partial}{\partial\alpha}$ \cite{Manton1985,IsoMurayama1990}. It is also easy to see that under small gauge transformations the wave functions \eqref{eqApp:WaveFuncitons} transform as $\psi_{n}(x)\rightarrow e^{ie\lambda(x)-ie\lambda(0)}\psi_{n}(x)$ and thus $c_{\sigma,n}\rightarrow e^{ie\lambda(0)}c_{\sigma,n}$. It is clear that the currents and its momentum modes, including the charges are invariant under all gauge transformations.

The homotopy grout of $U(1)$ symmetry is $\pi_{1}\left(U(1)\right)=\mathbb{Z}$ and large gauge transformations are generated by
\begin{equation}
\lambda(x)=\frac{2\pi}{eL}wx,\hspace{1cm}w\in\mathbb{Z}
\end{equation}
The wave functions $\psi_{n}(x)$ are invariant under those thus the fermion operators transform as $c_{n}\rightarrow c_{\sigma,n+w}$. Consequently, the $N$ vacua transform as $\left|0;N\right\rangle \rightarrow\left|0;N+w\right\rangle $. The large gauge transformations commute with the Hamiltonian so they can be diagonalised in the same basis. The eigenstates of large gauge transformations are the $\theta$ vacua
\begin{equation}
\left|\theta\right\rangle =\sum_{N\in\mathbb{Z}}e^{-iN\theta}\left|0;N\right\rangle,\hspace{1cm}\theta\in[0,2\pi)
\end{equation}
which form a continuous degenerate family of ground states of the fermionic parts of the Schwinger model Hamiltonian. A ground state of the full Hamiltonian is obtained as a tensor product of the ground state of the EM part with a $\theta$ vacuum.

\paragraph{\textbf{Hamiltonian. - }} Following from the gauge invariance constraints \eqref{eqApp:GaugeInvariance} we have 
\begin{equation}
H_{\text{EM}}=-\frac{1}{2L}\left[\left(\frac{\partial}{\partial\alpha}\right)^{2}+2\left(\frac{eL}{2\pi}\right)^{2}\sum_{n>0}\frac{1}{n}\left(b_{-,n}^{\dagger}-b_{+,n}\right)\left(b_{+,n}^{\dagger}-b_{-,n}\right)\right].
\end{equation}
Taking into account that $\left[H_{\text{F}},b_{\sigma,n}^{\dagger}\right]=\frac{2\pi}{L}nb_{\sigma,n}^{\dagger}$ and deducing its zero mode content from \eqref{eqApp:HexpOnN}, the Hamiltonian $H_{\text{F}}$ can only take the form
\begin{equation}
H_{\text{F}}=\frac{2\pi}{L}\sum_{\sigma=\pm}\left[\frac{1}{2}Q_{\sigma}^{2}-\frac{1}{24}+\sum_{n>0}nb_{\sigma,n}^{\dagger}b_{\sigma,n}\right].
\end{equation}
We can then split the massless part of the Schwinger model Hamiltonian into a part with zero modes and a part with nonzero momentum modes:
\begin{align}
H_{\text{EM}}+H_{\text{F}}&=H_{0}+\sum_{n>0}H_{n}-\frac{2\pi}{12L}\nonumber\\
H_{0}&=\frac{2\pi}{L}\left(\frac{Q^{2}+Q_{5}^{2}}{4}\right)-\frac{1}{2L}\left(\frac{\partial}{\partial\alpha}\right)^{2}\nonumber\\
H_{n}&=\frac{2\pi}{L}n\left(b_{+,n}^{\dagger}b_{+,n}+b_{-,n}^{\dagger}b_{-,n}\right)-\frac{e^{2}L}{4\pi^{2}n}\left(b_{+,n}^{\dagger}-b_{-,n}\right)\left(b_{-,n}^{\dagger}-b_{+,n}\right).
\end{align}
with $Q_5=Q_{+}-Q_{-}$ which takes the value $Q_5=2N-\frac{ecL}{\pi}+\delta_{b}-1$ on physical states and we keep in mind that $Q=0$ on physical states.

The zero mode Hamiltonian $H_{0}$ is the Hamiltonian of a massive harmonic oscillator with mass
\begin{equation}
M=\frac{e}{\sqrt{\pi}}
\end{equation}
and can be written in the canonical form  as 
\begin{equation}
H_0=M\left(B_{0}^{\dagger}B_{0}+\frac{1}{2}\right)\label{eqApp:ZeroModeHamiltonian}
\end{equation}
with $B_0=\sqrt{\frac{1}{2ML}}\left(-\sqrt{\pi}Q_{5}+\frac{\partial}{\partial\alpha}\right)$, $B_0^\dagger=\sqrt{\frac{1}{2ML}}\left(-\sqrt{\pi}Q_{5}-\frac{\partial}{\partial\alpha}\right)$.

The nonzero momentum Hamiltonians $H_{n}$ can be diagonalised with a Bogoliubov transformation
\begin{align}
B_{\sigma,n}&=\cosh(t_n) b_{\sigma,n} + \sinh(t_n) b_{-\sigma,n}^{\dagger}\nonumber\\
\cosh(t_{n})&=\frac{1}{2}\left(\frac{\sqrt{E_{n}}}{\sqrt{k_{n}}}+\frac{\sqrt{k_{n}}}{\sqrt{E_{n}}}\right)\nonumber\\
\sinh(t_{n})&=-\frac{1}{2}\left(\frac{\sqrt{E_{n}}}{\sqrt{k_{n}}}-\frac{\sqrt{k_{n}}}{\sqrt{E_{n}}}\right)\label{eqApp:Bogoliubov}
\end{align}
with $k_n=\frac{2\pi n}{L}$ and $E_n=\sqrt{M^2+k_n^2}$. Then
\begin{equation}
H_{n}=E_{n}\left(B_{+,n}^{\dagger}B_{+,n}+B_{-,n}^{\dagger}B_{-,n}+1\right)
\end{equation}
and $H_{\text{EM}}+H_{\text{F}}$ becomes the Hamiltonian of the free massive boson with the mass $M$. This reproduces the Schwinger's result that the QED in $D=1+1$ is gaped even if the bare mass of the fermion is zero. The Bogoliubov operator of this  transformation, $U_n b_{\sigma,n} U_n^\dagger = B_{\sigma,n}$, is the squeezing operator $U_n=\exp\left[-t_n\left(B_{+,n}^\dagger B_{-,n}^\dagger-B_{+,n}B_{-,n}\right)\right]$ meaning that the vacua annihilated by the massive modes $B_{\sigma,n}$ are the squeezed coherent $\theta$ vacua
\begin{equation}
\left|\theta\right\rangle_{M}=\left(\prod_{n>0}U_n\right)\left|\theta\right\rangle.
\end{equation}

It remains to treat the mass term in the Hamiltonian, $H_m$. We can express it in terms of the bosonic momentum modes of the currents, $b_{\sigma,n}$ using the relation
\begin{equation}
\Psi_{\sigma}(x)=F_{\sigma}\frac{1}{\sqrt{L}}:\negmedspace e^{-\sigma i \left(\sqrt{4\pi}\Phi_{\sigma}(x)-\frac{\pi}{L}\delta_b x \right)}\negmedspace:\label{eqApp:Bosonisation}
\end{equation}
with 
\begin{equation}
\Phi_{\sigma}(x)=\frac{1}{\sqrt{4\pi}}\left\{ \frac{2\pi}{L}Q_{\sigma}x-i\sum_{n>0}\frac{1}{\sqrt{n}}\left(b_{\sigma,n}e^{-\sigma in\frac{2\pi}{L}x}-b_{\sigma,n}^{\dagger}e^{\sigma in\frac{2\pi}{L}x}\right)+\sigma e\int_{0}^{x}dx'\,A(x')\right\} \label{eqApp:PhiField}
\end{equation}
and with the normal ordering with respect to the modes $b_{\sigma,n}$, which is the \textit{bosonisation relation} for a fermion coupled to the EM field. Here, the term with $\delta_b$ is to assure that the fermion field satisfies the correct boundary conditions and $F_{\sigma}$ are the Klein factors satisfying 
\begin{align}
\left[F_\sigma, A_{m}\right]=\left[F_\sigma,\frac{\delta}{\delta A_{m}}\right]=0,&\hspace{1.5cm}\left[F_\sigma,b_{\rho,m}\right]=\left[F_\sigma,b_{\rho,m}^{\dagger}\right]=0,\nonumber\\
\left[Q_\sigma,F_\rho^\dagger\right]=\delta_{\sigma,\rho}F_\rho^\dagger,&\hspace{1.5cm}\left[Q_\sigma,F_\rho\right]=-\delta_{\sigma,\rho}F_\rho,\nonumber\\
\left\{F_\sigma^\dagger,F_\rho\right\}=2\delta_{\sigma,\rho},&\hspace{1.5cm}F_\sigma^\dagger F_\sigma=1.\label{eqApp:KleinAlgebra}
\end{align}
Since a function of $b_{\sigma,n}$ and $b_{\sigma,n}^\dagger$ can never alter the fermion number, the Klein factors make sure that $\Psi_{\sigma}(x)$ as defined above has the true fermionic character. Some authors prefer to use exponentials of the zero modes of the compactified massless boson field in place of the Klein factors and the two conventions are fully equivalent. In particular, it can be shown that $\left\{\Psi_{\sigma}(x),\Psi_{\rho}(x)\right\}=\delta_{\sigma,\rho}\delta(x-y)$. Using the relations $\left[\frac{\delta}{\delta A_{n}},b_{+,n}^{\dagger}\right]=\left[\frac{\delta}{\delta A_{-n}},b_{+,n}\right]=\left[\frac{\delta}{\delta A_{-n}},b_{-,n}^{\dagger}\right]=\left[\frac{\delta}{\delta A_{n}},b_{-,n}\right]=\frac{eL}{2\pi\sqrt{n}}$ which follow from \eqref{eqApp:GaugeInvariance} it's easy to see that $\left[\frac{\delta}{\delta A(x)},\Psi_{\sigma}(y)\right]=0$. Finally, considering that $F_{\sigma}\rightarrow e^{ie\lambda(0)}F_{\sigma}$ under gauge transformations, if follows that $\Psi_{\sigma}(x)$ transforms as the fermion field. We also have, as follows from the second line of \eqref{eqApp:KleinAlgebra} that $F_{\sigma}^{\dagger}F_{-\sigma}\left|0;N\right\rangle  =\left|0;N+\sigma\right\rangle$ and thus:
\begin{equation}
F_{\sigma}^{\dagger}F_{-\sigma}\left|\theta\right\rangle_{M}=e^{\sigma i\theta}\left|\theta\right\rangle_{M}.\label{eqApp:KleinThetaExpectation}
\end{equation}
Using the definition \eqref{eqApp:PhiField} we can also read out the \textit{fermionisation relation} for the fermion field coupled to the EM field, the inverse of the bosonisation relation:
\begin{equation}
\partial_{x}\Phi_{\sigma}(x)=\sqrt{\pi}J_{\sigma}(x)+\frac{\sigma e}{2\sqrt{\pi}}\,A(x).
\end{equation}

We can use the bosonisation relation \eqref{eqApp:Bosonisation} to express the mass term in the Hamiltonian as
\begin{equation}
H_m=-m\frac{1}{L}e^{\sum_{n>0}\frac{1}{n}\left(1-\frac{k_{n}}{E_{n}}\right)}\int_{0}^{L}dx\,\sum_{\sigma=\pm} e^{\sigma i\frac{2\pi}{L}(1-\delta_b)x}:\negmedspace e^{\sigma i\sqrt{4\pi}\Phi(x)}\negmedspace:_{M}F_{\sigma}^{\dagger}F_{-\sigma},
\end{equation}
where we have defined $\Phi(x)\equiv \Phi_+(x) + \Phi_-(x)$ and $:\negmedspace\bullet\negmedspace:_{M}$ denotes normal ordering with respect to the massive modes $B_{\sigma,n}$. The prefactor $e^{\sigma i\frac{2\pi}{L}x}$ comes from commuting $F_{-\sigma}$ past $e^{i\sigma \frac{2\pi}{L}Q_{-\sigma}x}$ using \eqref{eqApp:KleinAlgebra}. The prefactor $e^{\sum_{n>0}\frac{1}{n}\left(1-\frac{k_{n}}{E_{n}}\right)}$ comes from  substituting the Bogoliubov transform \eqref{eqApp:Bogoliubov} into $\Phi_{\sigma}(x)$ and then rearranging the expression for $H_m$ into the normal ordered form w.r.t. $M$. In the $L\rightarrow\infty$ limit these prefactors take the value $\frac{ML}{4\pi}e^{\gamma}$ where $\gamma=0.5772\ldots$ is the Euler-Mascheroni constant.

Finally, putting all the terms together, with the $L\rightarrow\infty$ expression for the prefactor in the mass term, the Schwinger model Hamiltonian takes the form
\begin{align}
H&=M\left(B_{0}^{\dagger}B_{0}\right)+\sum_{n>0}E_{n}\left(B_{+,n}^{\dagger}B_{+,n}+B_{-,n}^{\dagger}B_{-,n}\right)+\text{const}
\nonumber\\
&\hspace{4cm}
-\frac{mM}{4\pi}e^{\gamma}\int_{0}^{L}dx\,\sum_{\sigma=\pm} e^{\sigma i\frac{2\pi}{L}(1-\delta_b)x} :\negmedspace e^{\sigma i\sqrt{4\pi}\Phi(x)}\negmedspace:_{M}F_{\sigma}^{\dagger}F_{-\sigma}\nonumber\\
&"\negmedspace=\negmedspace"\,\int_{0}^{L}\left[\frac{1}{2}\left(\Pi^2+(\partial_x\Phi)^2+M^2\Phi^2\right)-\frac{mM}{2\pi}e^{\gamma}:\negmedspace\cos\left(\sqrt{4\pi}\Phi(x)+\theta+\frac{2\pi}{L}(1-\delta_b)x\right)\negmedspace:_{M}\right]\label{eqApp:SchwingerHamiltonian}
\end{align}
where $\text{const}=\sum_{n>0}E_{n}+\frac{1}{2}M$ only affects the ground state energy and will be irrelevant to us. The last "equality" is to be understood only up to the details of the modes captured through the above bosonisation procedure and has taken into account \eqref{eqApp:KleinThetaExpectation} and the fact that all the physical states are created on top of the $\theta$ vacuum. The parameter $\theta$ thus appears in the Hamiltonian and plays the role of the constant background electric field as first pointed out by Coleman \cite{Coleman1975,Coleman1976}. As is manifest in the first line, the zero mode $B_{0}$ does not enter in the cosine term and is a harmonic oscillator decoupled from the other degrees of freedom. For our THM implementation, we choose $\delta_b=1$, the anti-periodic boundary conditions for the fermion, the Neveu-Schwarz sector.

\paragraph{\textbf{Hilbert space. - }}
As has been made explicit in the above discussion, the Hilbert space of the Schwinger model after eliminating the gauge redundancy takes the form of the tensor product of the Hilbert space of the zero modes with the Hilbert space generated by all the possible bosonic excitations on top of the theta vacuum. All together we can write any state in the Hilbert space in the form
\begin{equation}
\left|\vec{r}\right\rangle \equiv\frac{1}{N_{\vec{r}}}\left(B_{0}^{\dagger}\right)^{r_{0}}\prod_{n=1}^{\infty}\left(B_{-,n}^{\dagger}\right)^{r_{-,n}}\left(B_{+,n}^{\dagger}\right)^{r_{+,n}} \left|0\right\rangle_0\otimes\left|\theta\right\rangle_{M}\label{eqApp:BasisState}
\end{equation}
where $\vec{r}\equiv (r_{0},r_{-,1},r_{-,2},\ldots,r_{+,1},r_{+,2},\ldots)$ is a vector of occupation numbers and $\left|0\right\rangle_0$ is the vacuum of the $B_0$ mode. The normalisation is
$
N_{\vec{r}}^{2}=(r_{0}!)\prod_{k=1}^{\infty}(r_{k,-}!)(r_{k,+}!)
$.

\subsection{Truncated Hamiltonian method}

The truncated Hamiltonian method (THM) consists of splitting the Hamiltonian into an analytically
solvable and an unsolvable part, the perturbing potential. Then, expressing the
perturbing operator and the observables  as matrices in the eigenbasis of
the solvable part. Finally, an energy
cutoff is introduced which renders the matrices finite and enables
numerical diagonalisation which is the key to nonperturbative
treatment of a strong interaction with the THM. The above procedure of eliminating the redundant degrees of freedom and bosonising the model suggests a natural splitting of the Hamiltonian \eqref{eqApp:SchwingerHamiltonian} into the quadratic part $H_{\text{EM}}+H_{\text{F}}$ and the cosine potential $H_m$. 

In the following we first list the matrix elements in the Hilbert space of the quadratic part of the Hamiltonian for of all the required operators and then discuss how to implement the THM.

\paragraph{\textbf{Matrix elements. - }}

The matrix elements
are computed between general states of the Hilbert space $\left|\vec{r}\right\rangle $
and $\left|\vec{r}'\right\rangle $, defined in eq. (\ref{eqApp:BasisState})).
The required matrix elements are:

\uline{Boson mode operators:}
\begin{align}
\left<\vec{r}'\right|B_{\sigma,n}^{\dagger}\left|\vec{r}\right> & =\left(\prod_{\rho,k\neq \sigma, n}\delta_{r'_{\rho,k},r_{\rho,k}}\right)\sqrt{(r_{\sigma,n}+1)}\,\delta_{r'_{\sigma,n}-1,r_{\sigma,n}}\label{eqApp:BdagExp}\\
\left<\vec{r}'\right|B_{\sigma,n}\left|\vec{r}\right> & =\left(\prod_{\rho,k\neq \sigma, n}\delta_{r'_{\rho,k},r_{\rho,k}}\right)\sqrt{r_{\sigma,n}}\,\delta_{r'_{\sigma,n}+1,r_{\sigma,n}}\label{eqApp:Bexp}
\end{align}

\uline{Boson number operator:}
\begin{align}
\left<\vec{r}'\right|B_{\sigma,n}^{\dagger}B_{\sigma,n}\left|\vec{r}\right> & =r_{\sigma,n}\delta_{\vec{r}',\vec{r}}\label{eq:BdagBexp}
\end{align}

\uline{Vertex operator:}
To implement the cosine potential we need the matrix elements
\begin{align}
&\left<\vec{r}'\right|:\negmedspace e^{\rho i\sqrt{4\pi}\Phi(x)}\negmedspace:_{M}F_{\rho}^{\dagger}F_{-\rho}\left|\vec{r}\right> =\nonumber \\
& =e^{\rho i\theta}\left\langle \vec{r}'\right|\prod_{\sigma=\pm}\prod_{n=1}^{\infty}e^{-\rho\sqrt{\frac{2\pi}{L}}\frac{1}{\sqrt{E_{n}}}B_{\sigma,n}^{\dagger}e^{i\sigma k_n x}}e^{\rho\sqrt{\frac{2\pi}{L}}\frac{1}{\sqrt{E_{n}}}B_{\sigma,n}e^{-i\sigma k_n x}}\left|\vec{r}\right\rangle \nonumber \\
& =e^{\rho i\theta}\delta_{r'_0,r_0}\prod_{\sigma=\pm}\prod_{n=1}^{\infty}\frac{1}{\sqrt{r'_{\sigma,n}!r_{\sigma,n}!}}e^{i\sigma k_n x(r'_{\sigma,n}-r_{\sigma,n})}\cdot\nonumber \\
& \hspace{2cm}\cdot\sum_{j_{\sigma,n}'=0}^{\infty}\sum_{j_{\sigma,n}=0}^{\infty}\frac{(-1)^{j'_{\sigma,n}}}{j_{\sigma,n}!j'_{\sigma,n}!}\left(\sqrt{\frac{2\pi}{L}}\frac{\rho}{\sqrt{E_{k}}}\right)^{j_{\sigma,n}+j'_{\sigma,n}}\left\langle \left(B_{\sigma,n}\right)^{r'_{\sigma,n}}\left(B_{\sigma,n}^{\dagger}\right)^{j'_{\sigma,n}}\left(B_{\sigma,n}\right)^{j_{\sigma,n}}\left(B_{\sigma,n}^{\dagger}\right)^{r_{\sigma,n}}\right\rangle \label{eqApp:VertexMatrixElement}
\end{align}
with
\begin{align}
\left\langle \left(B_{\sigma,n}\right)^{r'_{\sigma,n}}\left(B_{\sigma,n}^{\dagger}\right)^{j'_{\sigma,n}}\left(B_{\sigma,n}\right)^{j_{\sigma,n}}\left(B_{\sigma,n}^{\dagger}\right)^{r_{\sigma,n}}\right\rangle  & =\left(\begin{array}{c}
r'_{\sigma,n}\\
j'_{\sigma,n}
\end{array}\right)\left(\begin{array}{c}
r_{\sigma,n}\\
j_{\sigma,n}
\end{array}\right)j'_{\sigma,n}!j_{\sigma,n}!(r_{\sigma,n}-j_{\sigma,n})!\delta_{r'_{\sigma,n}-j'_{\sigma,n},r_{\sigma,n}-j_{\sigma,n}}\Theta(r_{\sigma,n}\geq j_{\sigma,n})\label{eqApp:ExpectationBoson}
\end{align}
In the first line we have substituted in the Bogoliubov transformation \eqref{eqApp:Bogoliubov} and used \eqref{eqApp:KleinThetaExpectation} to evaluate the expectation value of the Klein factors. The last equality follows by a power expansion.
Upon the integration $\int_{0}^{L}dx\,\left<\vec{r}'\right|:\negmedspace e^{\rho i\sqrt{4\pi}\Phi(x)}\negmedspace:_{M}F_{\rho}^{\dagger}F_{-\rho}\left|\vec{r}\right>$,
the factor $\prod_{\sigma=\pm}\prod_{n=1}^{\infty}e^{i\sigma k_n x(r'_{\sigma,n}-r_{\sigma,n})}$ gives
the momentum conservation 
$\delta\left(\sum_{\sigma=\pm}\sigma\sum_{n=1}^{\infty}n(r'_{\sigma,n}-r_{\sigma,n})\right)$.
This is a manifestation of translation invariance and means that we
can diagonalise different total momentum sectors separately and compute
the dynamics only in the sector where the initial state resides, the total momentum zero sector, $\sum_{\sigma=\pm}\sigma\sum_{n=1}^{\infty}n r_{\sigma,n}=0$.
The expression for the matrix elements of the vertex operator
is a product of terms corresponding to the two chiralities which
is another property that facilitates the implementation. Furthermore,
it is clear that the vertex operator does not mix different sectors of the $B_0$ mode, so that the Hamiltonian can be diagonalised in each sector separately.

\uline{Observables:}

In the zero sector of the total momentum where the quench dynamics resides,
only those quadratic terms of bosonic modes in $C_{\mu}(t,x,y)$ give
nonzero contributions which preserve the momentum. Thus, the expectation
values are:
\begin{align}
\left\langle \Psi\right|:\negmedspace J^{0}(x)J^{0}(y) \negmedspace:\left|\Psi\right\rangle  & =\frac{1}{\pi L}\sum_{n=1}^{\infty}\frac{k_n^{2}}{E_{n}}\cos\left(k_n(x-y)\right)\left(\sum_{\sigma=\pm}\left\langle B_{\sigma,n}^{\dagger}B_{\sigma,n}\right\rangle _{\Psi}-\left\langle B_{-,n}B_{+,n}\right\rangle _{\Psi}-\left\langle B_{-,n}^{\dagger}B_{+,n}^{\dagger}\right\rangle _{\Psi}\right)\nonumber\\
\left\langle \Psi\right|:\negmedspace J^{1}(x)J^{1}(y)\negmedspace:\left|\Psi\right\rangle  & =\frac{1}{\pi L}\sum_{n=1}^{\infty}E_{k}\cos\left(k_n(x-y)\right)\left(\sum_{\sigma=\pm}\left\langle B_{\sigma,n}^{\dagger}B_{\sigma,n}\right\rangle _{\Psi}+\left\langle B_{-,n}B_{+,n}\right\rangle _{\Psi}+\left\langle B_{-,n}^{\dagger}B_{+,n}^{\dagger}\right\rangle _{\Psi}\right)
\end{align}
where we used the mode expansion of the currents \eqref{eqApp:CurrentExpansion}, expressed the charges in terms of the $B_0$ modes using the equations below \eqref{eqApp:ZeroModeHamiltonian}, abbreviated $\left\langle \bullet\right\rangle _{\Psi}\equiv\left\langle \Psi\right|\bullet\left|\Psi\right\rangle $
and dropped the diverging $\sum_{n=1}^{\infty}\frac{k_n^{2}}{E_{n}}\cos\left(k_n(x-y)\right)$
and $\sum_{n=0}^{\infty}E_{n}\cos\left(k_n(x-y)\right)$ by normal
ordering. The expectation values of the quadratic terms on a state
can be computed using the matrix elements (\ref{eqApp:BdagExp}),
(\ref{eqApp:Bexp}) and (\ref{eq:BdagBexp}).

To study the cluster violation of the correlators of chiral fermion fields $\left\langle\psi_\sigma^\dagger(x)\psi_{-\sigma}(x)\psi_{-\sigma}^\dagger(y)\psi_\sigma(y) \right\rangle$, one can use:
\begin{itemize}
	\item To get $\left\langle\psi_\sigma^\dagger(x)\psi_{-\sigma}(x)\right\rangle$, we bosonise using \eqref{eqApp:Bosonisation}, substitute the Bogoliubov transformed operators \eqref{eqApp:Bogoliubov} and normal order with respect to the massive modes:
	\begin{eqnarray}
	\psi_\sigma^\dagger(x)\psi_{-\sigma}(x)=\frac{1}{L}e^{\sum_{n>0}\frac{1}{n}\left(1-\frac{k_{n}}{E_{n}}\right)}e^{\sigma i\frac{2\pi}{L}(1-\delta_b)x}:\negmedspace e^{\sigma i\sqrt{4\pi}\Phi(x)}\negmedspace:_{M}F_{\sigma}^{\dagger}F_{-\sigma}\label{eqApp:BosonisedTwoPoint}
	\end{eqnarray}
	The matrix elements are given by eq. \eqref{eqApp:VertexMatrixElement} and notice that in the total momentum zero sector, the factor $\prod_{\sigma=\pm}\prod_{n=1}^{\infty}e^{i\sigma k_n x(r'_{\sigma,n}-r_{\sigma,n})}$ becomes just the identity, reflecting the translation invariance. 
	\item To get $\left\langle\psi_\sigma^\dagger(x)\psi_{-\sigma}(x)\psi_{-\sigma}^\dagger(y)\psi_\sigma(y) \right\rangle$, we use eq. \eqref{eqApp:BosonisedTwoPoint} and its conjugate with the Klein operator algebra \eqref{eqApp:KleinAlgebra}. Normal ordering the whole expression gives:
	\begin{equation}
	\psi_\sigma^\dagger(x)\psi_{-\sigma}(x)\psi_{-\sigma}^\dagger(y)\psi_\sigma(y)=\frac{1}{L^2}\left(e^{\sum_{n>0}\frac{1}{n}\left(1-\frac{k_{n}}{E_{n}}\right)}\right)^{2}e^{\frac{2\pi}{L}\sum_{n>0}\frac{2}{E_{n}}\cos\left(\frac{2\pi}{L}n(x-y)\right)}:\negmedspace e^{\sigma i\sqrt{4\pi}\left(\Phi(x)-\Phi(y)\right)}\negmedspace:_{M}
	\end{equation}
	Recall that  $\lim\limits_{L\rightarrow\infty}e^{\sum_{n>0}\frac{1}{n}\left(1-\frac{k_{n}}{E_{n}}\right)}=\frac{ML}{4\pi}e^{\gamma}$ where $\gamma$ is the Euler-Mascheroni constant so that the explicit $L$ dependence cancels out. In fact, we use this limiting expression to get the results closer to the thermodynamic limit. The required matrix element is given by
	\begin{align}
	&\left<\vec{r}'\right|:\negmedspace e^{\rho i\sqrt{4\pi}\left(\Phi(x)-\Phi(y)\right)}\negmedspace:_{M}\left|\vec{r}\right> =\nonumber \\
	& =\delta_{r'_0,r_0}\prod_{\sigma=\pm}\prod_{n=1}^{\infty}\frac{1}{\sqrt{r'_{\sigma,n}!r_{\sigma,n}!}}\sum_{j_{\sigma,n}'=0}^{\infty}\sum_{j_{\sigma,n}=0}^{\infty}\frac{(-1)^{j'_{\sigma,n}}}{j_{\sigma,n}!j'_{\sigma,n}!}\left(\sqrt{\frac{2\pi}{L}}\frac{\rho}{\sqrt{E_{k}}}\right)^{j_{\sigma,n}+j'_{\sigma,n}}\cdot\nonumber \\
	& \hspace{2cm}\cdot\left(e^{i\sigma k_{n}x}-e^{i\sigma k_{n}y}\right)^{j'_{\sigma,n}}\left(e^{-i\sigma k_{n}x}-e^{-i\sigma k_{n}y}\right)^{j_{\sigma,n}}\left\langle \left(B_{\sigma,n}\right)^{r'_{\sigma,n}}\left(B_{\sigma,n}^{\dagger}\right)^{j'_{\sigma,n}}\left(B_{\sigma,n}\right)^{j_{\sigma,n}}\left(B_{\sigma,n}^{\dagger}\right)^{r_{\sigma,n}}\right\rangle 
	\end{align}
	and the expectation value of the boson operators is given by eq. \eqref{eqApp:ExpectationBoson}. 
	
\end{itemize}

\paragraph{\textbf{Truncation. - }}

We preform the THM truncation by choosing a value for the cutoff energy
$E_{\text{cut}}$ and keeping only those states of the Hilbert space
$\left|\vec{r}\right\rangle $ for which $\left\langle \vec{r}\right|H_{\text{EM}}+H_{\text{F}}\left|\vec{r}\right\rangle \leq E_{\text{cut}}$.
This results in a better converging code than for example if truncating
by keeping a fixed number of momentum modes. The truncation criterium depends on the charge
$e$ and the system size $L$ (as $E_{n}=\sqrt{k_n^{2}+\frac{e^{2}}{\pi}}=\frac{2\pi}{L}\sqrt{n^{2}+\frac{L^{2}}{(2\pi)^{2}}\frac{e^{2}}{\pi}}$)
and for fixed $E_{\text{cut}}$ the number of states in the THM Hilbert
space decreases with increasing $e$ and $L$. Therefore, in practice
the truncation is done by choosing a desired number
of states in the THM Hilbert space and then for a given $e$ and $L$
finding $E_{\text{cut}}$ that gives us a Hilbert space size closest
to the desired one. In that way we can assure that results obtained
at different $e$ and $L$ are achieved with comparable Hilbert space
sizes.

The size of the Hilbert space that has to be kept in the computer's memory can be reduced by taking into account the symmetries of the model. Since the zero mode $B_0$ is decoupled from the rest of the modes, we can diagonalise the Hamiltonian in each of it's sectors separately. In particular, for real time dynamics following quenches it is enough to keep the $\left\langle B_0^\dagger B_0\right\rangle=0$ sector where the initial states, the ground states, reside.
Furthermore, because of the translation invariance of the model, the ground
states are in the the zero total momentum sector ($p_{\text{tot}}=\sum_{\sigma=\pm}\sigma\sum_{n=1}^{\infty}k_n\left\langle B_{\sigma,n}^{\dagger}B_{\sigma,n}\right\rangle = 0$)
of the Hilbert space, which drastically reduces the number of states
that have to be kept in the computer's memory in order to compute
the quench dynamics. We do, however, have to diagonalise the Hamiltonian also in the sectors
with other values of the total momentum in order to compute the full
spectrum of the model (excited states). For the results presented in this Letter, we use up to 20 000 states per sector.

In case of truncated conformal space approach (TCSA) methods, where the expansion is around a CFT, the renormalisation
group theory guarantees that for relevant perturbing operators, the
cut-away high energy part of the Hilbert space is only very weakly
coupled to the low energy part and therefore does not modify the low
energy physics that one studies with such methods \cite{Rychkov2017,KonikReviewNonperturbative2018}.
In a more general expansion like we use here, we cannot directly rely
on the RG theory and have to establish convergence by extensive tests.
We have therefore tested that all our results have converged with
the THM cutoff. We have also tested that the scalar particle mass
computed with our method agrees with matrix product states (MPS) and
tensor network (TN) computations with a discretised version of the
Schwinger model \cite{Banuls2013,Buyens2015,BuyensThesis2016} (fig.
2 in the main text) and that in the $e\rightarrow0$ limit we recover the spectrum of the sine-Gordon model.

\paragraph{\textbf{Quench protocol. - }}

In order to study the quench dynamics, one takes for the initial state
the ground state $\left|\Psi\right\rangle $ of the prequench Hamiltonian
$H(m_{0}/e_{0},\theta_{0},L)$ which can be found
by numerical diagonalisation of the Hamiltonian. At $t=0$, the parameters
are quenched to the postquench values $H(m/e,\theta,L)$.
The dynamics is computed using the numerical exponentiation of the
postquench Hamiltonian:
\begin{equation}
\left|\Psi(t)\right\rangle =e^{-itH}\left|\Psi\right\rangle .
\end{equation}
Finally, correlators are computed as expectation values on these states
\begin{equation}
C_{\mu}(t,x,y)  =\left\langle J^{\mu}(t,x)J^{\mu}(t,y)\right\rangle.
\end{equation}

\section{Further results}

Here we list some further results adding more detail to those presented in fig. 3 in the main text.

\begin{figure*}[htbp]
	\begin{centering}
		\includegraphics[width=\textwidth]{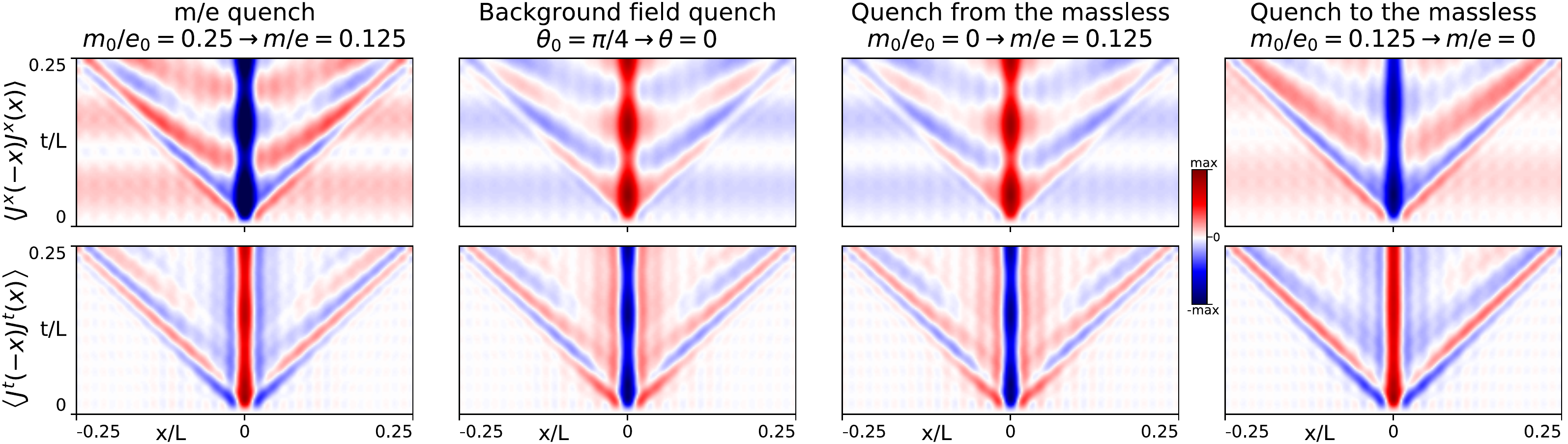}
		\par\end{centering}
	\caption{Time dependent $\left\langle J^{x}(t,x)J^{x}(t,y)\right\rangle $
		and $\left\langle J^{t}(t,x)J^{t}(t,y)\right\rangle $ correlations
		for different type of quenches in the Schwinger model (initial correlations subtracted): 1.) Quench in $m/e$ with $m_{0}=0.25$,
		$m=0.125$, $\theta_{0}=\theta=0$; 2.) Quench
		in $\theta$ with $\theta_{0}=\frac{\pi}{4}$, $\theta=0$, $m_{0}=m=0.125$; 3.) Quench from the massless Schwinger   model with $m_{0}=0$,
		$m=0.125$, $\theta_{0}=\theta=0$; 4.) Quench to the massless model $m_{0}=0$,
		$m=0.125$, $\theta_{0}=\theta=0$. All with $e_{0}=e=1$, $L=40$.
		\label{fig:SchwingerQuenchesSupplement}}
\end{figure*}

The effect is found in quenches of either of the parameters of the system, $e/m$ and $\theta$ as well as in quenches to and from the massless Schwinger model. The sign of the out-of-horizon correlations changes depending whether the quenched parameter is increased or decreased. Fig \ref{fig:SchwingerQuenchesSupplement} gives an overview of these observations.

\begin{figure*}[htbp]
	\begin{centering}
		\includegraphics[width=\textwidth]{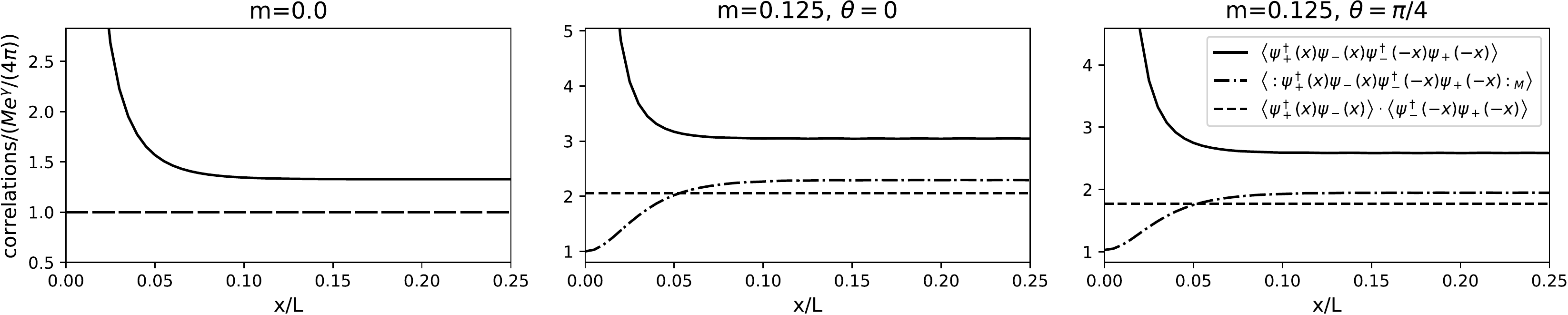}
		\par\end{centering}
	\caption{Cluster violation of the $\left\langle\psi_\sigma^\dagger(x)\psi_{-\sigma}(x)\psi_{-\sigma}^\dagger(y)\psi_{\sigma}(y)\right\rangle$ correlator at different values of the parameters: 1.) $m=0$, $\theta=0$; 2.) $m=0.125$, $\theta=0$; 3.) $m=0.125$, $\theta=\pi/4$. All with $e_{0}=e=1$, $L=40$.
		\label{fig:ClusterViolation}}
\end{figure*}

Fig. \ref{fig:ClusterViolation}: The correlator $\left\langle\psi_\sigma^\dagger(x)\psi_{-\sigma}(x)\psi_{-\sigma}^\dagger(y)\psi_{\sigma}(y)\right\rangle$ exhibits clustering. While the clustering is restored by normal ordering for the massless model, it is violated in the massive model even for the normal ordered correlator. The magnitude of the correlators depends on both $e/m$ and $\theta$.

\begin{figure*}[htbp]
	\begin{centering}
		\includegraphics[width=0.5\textwidth]{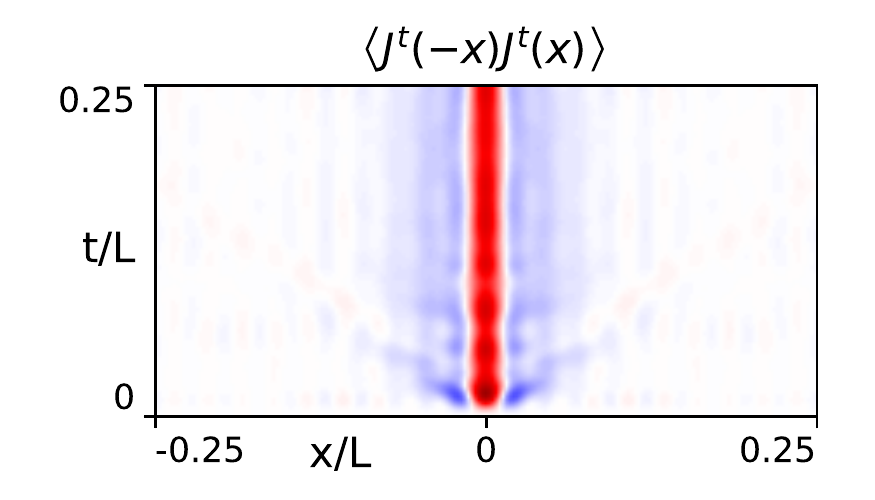}
		\par\end{centering}
	\caption{Suppression of the horizon spreading in quenches to the $\theta=\pi$ line above the Ising phase transition point. Here, $\theta_{0}=0$, $\theta=\pi$, $m_{0}=m=0.5615$, $e_{0}=e=1$, $L=40$.
		\label{fig:Confinement}}
\end{figure*}

Fig \ref{fig:Confinement}: In quenches to the special value of the parameter $\theta=\pi$, to the mass above the Ising transition point, the horizon dynamics is strongly suppressed, resembling the confined dynamics observed in \cite{TakacsConfinement2017}. Note that here we are plotting the $C_t$ correlator for which there is no horizon violation effect.

\end{document}